\documentclass[aps,prb,amsfonts,amsmath,amssymb,bm
,twocolumn
]{revtex4}
\usepackage{graphicx}
 
\begin{document}
\flushbottom
\title{Low-temperature structural phase transitions: \\Phonon-like and relaxation order-parameter dynamics}

\author{A. Cano}
\email{andres.cano@uam.es}
\author{A. P. Levanyuk}
\email{levanyuk@uam.es}
\affiliation{Departamento de F\'\i sica de la Materia Condensada C-III, Universidad Aut\'onoma de Madrid, E-28049 Madrid, Spain}

\date{\today}

\begin{abstract}
A theoretical study on low-temperature structural phase transitions is presented, in which both phonon-like and relaxation order-parameter dynamics are contemplated. While the first limiting case has been considered previously, the second one is studied here for the first time. Attention is put on the low-temperature asymptotics of the temperature dependence of the generalized susceptibility. In the relaxation case, it is found $\sim (T^2 - T_c^2)^{-1}$ for temperature-independent relaxation times in a broad region of the temperature-pressure phase diagram. In contrast to the obtained in the phonon-like case, this asymptotics is not modified by long-range interactions (dipolar forces, piezoelectric effect, etc.).
\end{abstract}


\maketitle
 
\section{Introduction}

Structural phase transitions are usually divided into two classes: displacive and order-disorder ones (see, e.g., Ref. \onlinecite{Strukov_Levanyuk}). The labels refer to limiting cases, but still the division is convenient. Among other features, it points out the order-parameter dynamics: phonon-like in the former and relaxational in the latter. By varying a control parameter, e.g. pressure or chemical composition, the corresponding transition temperature can virtually be driven to zero in both cases. Well known examples are SrTiO$_3$ (displacive)\cite{Uwe76} and KH$_2$PO$_4$ (order-disorder).\cite{especial_KDP} The theory of low-$T$ displacive phase transitions\cite{Barrett52,Rechester71,Khmelnitskii71,Schneider76} has been developed since long time ago (see Ref. \onlinecite{Kvyatkovskii01} for a recent review). However, the one concerning to low-$T$ order-disorder transitions is still lacking. 

It is worth mentioning the possibility that, depending on the temperature regime one considers, the order-parameter dynamics of a given transition may evolve from one of the above mentioned limiting cases to the another one. Let us consider, for instance, a relaxational one. With lowering the transition temperature this dynamics may convert into phonon-like because of, e.g., the increasing importance of tunneling.\cite{nota00} But quantum effects may be important before that such a conversion, if actually exists, takes place.\cite{nota0} Existing theories on low-$T$ structural phase transitions\cite{Barrett52,Rechester71,Khmelnitskii71,Schneider76,Kvyatkovskii01} are therefore clearly insufficient to interpret these cases, as far as they are restricted to phonon-like dynamics, i.e., they neglect any damping. The aim of this paper is just to account for this damping and, ultimately, to describe low-$T$ structural phase transitions with relaxational order-parameter dynamics.

The importance of the order-parameter dynamics in low-$T$ phase transitions is due to intimate connection existing between statistics and dynamics in quantum systems. This has been noticed, for instance, in magnets. Here various types of order-parameter dynamics have been revealed, and much attention has been paid to the role that these dynamics play in the corresponding low-$T$ transitions (see Ref. \onlinecite{Vojta03} for a recent review). As we have mentioned, different types of order-parameter dynamics are also possible in low-$T$ structural phase transitions. Here, in contrast, only the phonon-like one has been studied until now. In this paper, we shall study both limiting cases of phonon-like and relaxation order-parameter dynamics in structural phase transitions, paying attention to possible differences in the corresponding anomalies. To do so, we shall use a semiphenomenological approach in which low-$T$ phase transitions are described within a continuous media theory. Within this formalism, we shall assume that relaxation is characterized by a single phenomenological parameter (a viscosity coefficient) describing the simplest case of wavevector and frequency independent damping. To determine the microscopic nature of such a damping is a very difficult task that, being the matter of a full microscopic theory, is beyond the scope of the present work. So its temperature dependence, for instance, is assumed to be obtainable from the experiments. In the following we shall consider a temperature-independent damping for the sake of illustration, although this dependence would modify final formulas in a trivial way. 

It is worth mentioning that the lack of studies dealing with relaxational dynamics in low-$T$ structural phase transitions sometimes give rise to some confusion. In Ref. \onlinecite{Endo02}, for instance, it is reported the low-$T$ thermal anomaly of the dielectric constant occurring in prototypical order-disorder systems (KH$_2$PO$_4$ and KD$_2$PO$_4$). An order-parameter dynamics of relaxation type is well possible at these temperatures. Nevertheless, the authors analyze the experimental data on the basis of models in which this relaxation is completely absent.\cite{Tokunaga87} Their striking conclusion that, on the basis of corresponding discrepancies between theory and experiments, a classical order-disorder transition becomes displacive at low temperatures is therefore unjustified (see below).

In this work, we focus our attention in the anomalies of the generalized susceptibility (i.e. the dielectric susceptibility when considering ferroelectrics) associated with low-$T$ phase transitions because of the current experimental interest.\cite{Endo02} Nevertheless, it is provided a theoretical framework according to which all other anomalies can be obtained. The paper is organized as follows. Sec. II has a pedagogical character, and here we consider a simple model of one-component order-parameter in which long-range forces are neglected. It serves us to introduce the method to include relaxational dynamics when studying low-$T$ structural phase transitions. We first reproduce by this method the results of the existing (phononic) theory, and then extend this theory to relaxation case. We shall do it in two different ways. One of them consists in to calculate the response function from the nonlinear equation of motion (Sec.II.A), and the other one is by computing the free energy (Sec.II.B). In Sec. III we improve the model by accounting for acoustic phonons. In Sec. IV we shall focus our attention to KDP-type systems. In addition to considering here the relaxation case of the order-parameter dynamics, we revise the phonon-like case because we found that the correct low-temperature asymptotics of the temperature dependence of the susceptibility is different from that previously reported.

\section{Simple model}

Suppose that the phase transition at zero temperature takes place because of change in pressure. Integrating out all degrees of freedom but the critical ones, the effective potential energy per volume unit of the system can be written as
\begin{align}
U=\int\Big({a\over 2}\eta^2 + {b\over 4}\eta^4 + {c\over 2}(\nabla\eta)^2\Big) d{\mathbf r}
\label{U_model}\end{align}
(it is assumed that the system has unit volume). Here $\eta$ is the order parameter and $a=a_p(p-p_c)$ is the only pressure dependent coefficient. $p_c$ would correspond to the phase transition pressure in the classical case, i.e., if the ions would have infinite masses and therefore quantum fluctuations were suppressed. Real phase transition pressure is different and will be calculated below. 

Let us here make a comment relative to fluctuations. It is frequently said that, while usual (high-$T$) phase transitions are due to thermal fluctuations, that occurring at low temperatures are due to quantum ones. The term ``quantum phase transition'' seems then to be synonym of low-$T$ phase transition. This makes sense for some systems,\cite{Vojta03} but it is quite confusing for others. Among these others we have just the structural ones. In fact, as we have pointed out, a zero-$T$ structural phase transition is perfectly imaginable without any quantum fluctuation: a crystal with infinite-mass ions may change its structure by applying pressure in accordance with Eq. \eqref{U_model}. This observation does not mean, as we shall see below, that quantum fluctuations and/or quantum statistic play no role in low-$T$ structural phase transitions. 

\subsection{Generalized susceptibility}

Let us first calculate the static generalized susceptibility following a standard way and considering the classical case. The equation of motion for the order parameter can be written as  
\begin{align}
\rho_c \ddot \eta + \widetilde \gamma\dot \eta + a \eta + b \eta^3 -c\nabla^2\eta= h,
\label{lange_eta}\end{align}
where $\rho_c$ is an optic density and $\widetilde\gamma$ a ``viscosity coefficient.'' It is assumed that such a viscosity arises because of the coupling between the order parameter an other degrees of freedom not considered here explicitly. Consequently $\widetilde\gamma$ may depend, e.g., on temperature. But in the following we shall take it as a constant for the sake of simplicity. Thermal fluctuations are described by the random field $h$. In the classical case it is such that
\begin{align}
\langle h({\mathbf r},t) h({\mathbf r}', t')\rangle =
2\widetilde\gamma\beta^{-1}\delta({\mathbf r}-{\mathbf r'})\delta(t-t'),
\label{}\end{align}
where $\beta ^{-1}=k_B T$ and $\langle \dots \rangle$ denotes statistical average.

Let us express the order parameter as a Fourier transform: $\eta(\mathbf r,t) = \sum_{\mathbf k}\int d\omega\eta_{{\mathbf k},\omega} e^{i\mathbf k \cdot \mathbf r}e^{-i\omega t}$. Within a first approximation in the anharmonicity, the inverse of the static susceptibility associated with the macroscopic degree of freedom (${\mathbf k}=0$) can be taken as  
\begin{align}
\chi^{-1} = {a + 3b\sum_{\mathbf k}\int d\omega
\langle\eta_{{\mathbf k},\omega}\eta_{-{\mathbf k},-\omega}\rangle}.
\label{}\end{align}
The average in this expression can be computed by using the dynamic susceptibility in the harmonic case ($b=0$):
\begin{align}
\chi^{(0)}(k,\omega)&={1\over \rho_c[\omega_c^2(k)-\omega^2 -i\gamma \omega]}\nonumber\\
&=-{1\over \rho_c[\omega+i\lambda_1(k)][\omega+i\lambda_2(k)]},
\label{dyn_sus}\end{align}
where $\omega_c^2(k)=(a + ck^2)/\rho_c$, $\gamma = \widetilde\gamma /\rho_c$ and 
\begin{align}
\lambda_{1,2}(k)={1\over 2}\left\{
{\gamma}\pm i\left[4\omega_c^2(k)-{\gamma^2}\right]^{1/2}\right\}.
\label{lambda_12}\end{align}
We then have
\begin{align}
\chi^{-1}&= a+\int{d^3k\over (2\pi)^3} \int_0^\infty {d\omega} {2k_BT\over \pi\omega}\text{ Im }\chi^{(0)}(k,\omega).
\label{eq_chi_first}\end{align}

Although this analysis is strictly classical, it has been shown that the corresponding fully quantum mechanical expression follows merely with the replacement $k_BT \to (\hbar \omega /2)\coth (\hbar \beta \omega/2)$ (see, e.g. Refs. \onlinecite{Landau} and \onlinecite{Caldeira83}). We then have
\begin{widetext}
\begin{align}
\chi^{-1}&= a+ {\hbar\over \pi}\int{d^3k\over (2\pi)^3} \int_0^\infty d\omega \text{ Im }\chi^{(0)}(k,\omega)\coth (\hbar \beta \omega/2)
=a+{1\over \beta\rho_c}\int{d^3k\over (2\pi)^3}\sum_{n=-\infty}^{\infty} 
{1\over \nu_n^2 + |\nu_n|\gamma +\omega_c^2(k)}
\nonumber \\
&=a+{2\over \beta\rho_c}\int{d^3k\over (2\pi)^3}
\left({1\over 2\omega_c^2(k)}+{\psi[1+\lambda_1(k)/\nu]-\psi[1+\lambda_2(k)/\nu]\over \nu[\lambda_1(k)-\lambda_2(k)]}
\right),\label{eq_chi_first}\end{align}\end{widetext}
where $\nu_n=2\pi n/(\hbar \beta)$, $\nu = \nu_1$, and the psi function is the logarithmic derivate of the gamma function. [The infinite sum in the second form of Eq. \eqref{eq_chi_first} appears as a result of integration in the complex $\omega$-plane.]

\subsubsection{Phonon-like dynamics}

If there is no damping ($\gamma=0$) we have $\lambda_1=-\lambda_2=i\omega_c$. The resulting expression in Eq. \eqref{eq_chi_first} is 
\begin{align}
\chi^{-1}&= a^*_\text{ph}+ 
{3\hbar b\over 4\pi^2\rho_c}
\int
{n[\omega_c(k)]\over \omega_c(k)} k^2dk,
\label{suscep_simple_1}\end{align}
where 
\begin{align}
a^*_\text{ph} = a + {3\hbar b\over 8\pi ^2\rho_c}\int
{k^2dk\over \omega_c(k)},
\label{}\end{align}
and $n(\omega)= [\exp(\hbar\beta \omega)-1]^{-1}$ is the Bose-Einstein distribution function. Eq. \eqref{suscep_simple_1} coincides with the corresponding expression reported in Refs. \onlinecite{Rechester71} and \onlinecite{Khmelnitskii71}.

The transition pressure at zero temperature is such that $a^*_\text{ph}=0$, instead of $a=0$. This renormalization of the transition pressure is due to zero-point fluctuations. The critical frequencies $\omega_c(k)$ are also renormalizated in such a way, but one has to consider higher-order corrections to obtain it explicitly. Below we shall assume that $\omega_c(k)$ represents these renormalized frequencies. Thus it is reproduced the vanishing of the optical frequency $\omega_c(0)$ that takes place within this model as a result of the phase transition, i.e., the softening of the optical branch. 

Far from the transition point [$k_BT \ll \hbar\omega_c(0)$] the inverse of the susceptibility has an exponential dependence with temperature $\propto T^{3/2}\exp{[-\hbar \omega_c(0)/(k_BT)]}$. It is because of the ``inefficiency'' of the thermal activation of the optical phonons due to the hardness of this branch. The softening of the branch associated with the transition increases the ``efficiency'' of this thermal activation. Thus, close to the transition [$k_B T \gg \hbar\omega_c(0)$] the temperature dependence of the inverse of the susceptibility becomes $\propto T^2$. 

\subsubsection{Relaxational dynamics}

In the pure relaxation limit ($\rho_c \to 0$) we have $\lambda_1(k)\approx \varpi_c(k)$ and $\lambda_2(k)\approx \gamma -\varpi_c(k) $, where $\varpi_c(k) = \omega_c^2(k)/\gamma ={\cal K}(k)/\widetilde\gamma$ then describes the inverse of the relaxation times in the symmetric phase [see Eq. \eqref{lange_eta}]. In this case, the low-$T$ asymptotics of Eq. \eqref{eq_chi_first} can be obtained by using the asymptotic expansion of the psi function (it will be valid over the major of the integration interval). Thus we get 
\begin{align}
\chi^{-1} &
\approx a^*_\text{rel} + C T^2,
\label{chi_simple_asym}\end{align}
where
\begin{gather}
a^*_\text{rel}= a + 
{3\hbar b\over 2\pi^3 \gamma'}\int
k^2 {\ln{[\gamma/\varpi_c(k)]}}dk,\\
C=
{k_B^2 b\over 2\pi \hbar \gamma'}\int
{k^2dk\over \varpi_c^2(k)}.
\label{coefficient_c}\end{gather}

In the phonon case, the vicinity of the phase transition increases the thermal activation of phonons. This leads to different behaviors in the temperature dependence of the susceptibility depending on the distance from the transition point. In the relaxation case, however, the distance from the transition point is not so decisive for the thermal activation. In consequence, only the $\sim T^2$ behavior is obtained in the low-temperature regime of the relaxation case for a temperature-independent viscosity coefficient.

\subsection{Generalized susceptibility revisited: A method for further calculations}

Let us now reproduce the above results, but following a different method. The starting point is now the calculation of the Landau potential whatever the order-parameter dynamics be, i.e. the free energy as a function of the critical macroscopic degree of freedom. Once this Landau potential is obtained, all the anomalies associated with the low-$T$ structural phase transition under consideration can be computed. However, as we have mentioned, we shall restrict ourselves to calculate here the one of the generalized susceptibility. 

Let us now express the order parameter as a Fourier transform: $\eta(\mathbf r,t) = \sum_{\mathbf k}\eta_{\mathbf k}(t) e^{i\mathbf k \cdot \mathbf r}$. Thus, Eq. \eqref{U_model} can be rewritten showing explicitly its dependence on the above mentioned macroscopic degree of freedom $\eta_0$:
%\begin{widetext}
\begin{align}
U&=U_0(\eta_0)+\sum_{\mathbf k\not = 0}{{\cal K}(k,\eta_0)\over 2}|\eta_{\mathbf k}|^2 
\nonumber \\&\quad
+
b\eta_0\sum_{\mathbf k,\mathbf k' \not = 0}
\eta_{\mathbf k}\eta_{\mathbf k'}\eta_{-\mathbf k-\mathbf k'}
\nonumber \\
&
\quad+{b\over 4} \sum_{\mathbf k,\mathbf k',\mathbf k''\not = 0}
\eta_{\mathbf k}\eta_{\mathbf k'}\eta_{\mathbf k''}\eta_{-\mathbf k-\mathbf k'-\mathbf k''},
\label{U_Fourier}\end{align}
where
\begin{align}
U_0(\eta_0)&={a\over 2}\eta_0^2 + {b\over 4}\eta_0^4,\\ 
{\cal K}(k,\eta_0)&=a + 3b\eta_0^2 + ck^2.
\end{align}

The equations of motion for the Fourier components of the order parameter can be written as
\begin{align}
\ddot\eta_{\mathbf k}+\gamma\dot\eta_{\mathbf k}
+\omega_c^2(k,\eta_0) \eta_{\mathbf k}=0,
\label{motion_eta_std}\end{align}
where $\omega_c^2(k, \eta_0)={\cal K}(k,\eta_0)/\rho_c$. Here we have neglected the contribution given by the last two sums in Eq. \eqref{U_model}. Such a neglection is possible by virtue of a weak anharmonicity. But note that we do not neglect this anharmonicity at all: it is the responsible for the dependence of characteristic frequencies $\omega_c$ on $\eta_0$. The anharmonicity that we are partially neglecting [for instance, by omitting the last two terms in Eq. \eqref{U_Fourier}] may give some contribution to the relaxation which vanishes at zero temperature.\cite{Gurevich91} Accordingly $\gamma$ may depend on temperature and/or wavevector $\mathbf k$. But as before we shall take it as a constant.

\subsubsection{Phonon-like dynamics}

Let us present the method assuming first that the order-parameter dynamics is phonon-like. If the anharmonicity is weak, the equations of motion for the Fourier components of the order parameter can be written as
\begin{align}
\ddot\eta_{\mathbf k}
+\omega_c^2(k,\eta_0) \eta_{\mathbf k}=0.
\label{phono_motion}\end{align}
Within this approximation the system reduces to a system of decoupled harmonic oscillators. So its free energy can be written as\cite{Landau}
\begin{align}
\Phi(\eta_0)&= U_0(\eta_0)+
{1\over\beta}
\sum_{\mathbf k}\ln \left\{2\sinh \left[\hbar \beta\omega_c(k,\eta_0)/2\right]\right\}
\nonumber \\&= U_0(\eta_0)+
\sum_{\mathbf k}
{\hbar \omega_c(k,\eta_0)\over 2}
\nonumber \\&\quad +{1\over\beta}
\sum_{\mathbf k}
\ln \left\{
1- \exp \left[-\hbar \beta\omega_c(k,\eta_0)\right]\right\}.
\label{Landau_phonon}\end{align}
All degrees of freedom but that associated with $\eta_0$ are integrated out in this potential. However, it describes nonequilibrium states of the system just because, at this point, $\eta_0$ has a nonfixed value. Then one has to realize that what Eq. \eqref{Landau_phonon} represents is just the Landau potential of the system.\cite{Strukov_Levanyuk} 

We are now in a position to reproduce all the anomalies associated with the phase transition following well-known procedures.\cite{Strukov_Levanyuk} Let us consider the anomaly of the generalized susceptibility $\chi$. In the symmetric phase we have  
$\chi^{-1}=\Phi'' \equiv \left({\partial^2 \Phi/\partial \eta_0^2}\right)_{\eta_{0}=0}$ (hereafter prime denote partial differentiating with respect to $\eta_0$). Bearing in mind that
\begin{subequations}\begin{gather}
\omega_c'(k)=0, \\
\omega_c''(k)={3b/[\rho_c \omega_c(k)]},
\end{gather}\label{omega_c_primes}\end{subequations}
from Eq. \eqref{Landau_phonon} one finds that 
\begin{align}
\chi^{-1} &= a^*_\text{ph}+ 
{3\hbar b\over 4\pi^2\rho_c}
\int
{n[\omega_c(k)]\over \omega_c(k)} k^2dk,
\label{suscep_simple}\end{align}
where 
\begin{align}
a^*_\text{ph} = a + {3\hbar b\over 8\pi ^2\rho_c}\int
{k^2dk\over \omega_c(k)},
\label{}\end{align}
and $n(\omega)= [\exp(\hbar\beta \omega)-1]^{-1}$ is the Bose-Einstein distribution function. [Summation over wavevectors has been replaced by integration: $\sum_{\mathbf k}\approx (2\pi)^{-3}\int d{\mathbf k}$.] Eq. \eqref{suscep_simple} coincides with the one obtained in Sec.II.A.1. 

\subsubsection{Relaxational dynamics}

Let us now calculate the low-temperature asymptotics of the temperature dependence of the susceptibility when the motion of the order parameter includes some relaxation. Similar to the exposed in Sec. II. A, let us consider the equations
\begin{align}
 \ddot\eta_{\mathbf k}
+\gamma\dot\eta_{\mathbf k}
+\omega_c^2(k,\eta_0) \eta_{\mathbf k}=0.
\label{motion_eta}\end{align}
The system is therefore a system of decoupled damped harmonic oscillators. Making use of the partition function of such damped oscillators\cite{Weiss} the Landau potential can be obtained. It reads 
\begin{widetext}
\begin{align}
\Phi(\eta_0)&= U_0(\eta_0)+
{1\over\beta}
\sum_{\mathbf k}\left\{
\ln \hbar \beta \big[\lambda_1(k,\eta_0)\lambda_2(k,\eta_0)\big]^{1/2} 
-\ln\Gamma\big[1+\lambda_1(k,\eta_0)/\nu\big] \; -\ln\Gamma\big[1+\lambda_2(k,\eta_0)/\nu\big]
\right\},
\label{F_relax}\end{align}\end{widetext}
where $\Gamma $ is the gamma function and 
$\lambda_{1,2}(k,\eta_0)$ 
are given by Eq. \eqref{lambda_12} with $\omega_c(k)\to \omega_c(k,\eta_0)$. 

The inverse of the susceptibility in the symmetric phase can be calculated from Eq. \eqref{F_relax} by differentiating with respect to $\eta_0$. Noting that $\lambda_{1,2}(k,\eta_0)$ satisfy the relations 
\begin{subequations}\label{Vieta_2}\begin{gather}
\lambda_1(k) + \lambda_2(k) =\gamma,\\
\lambda_1(k)\lambda_2(k) =\omega_c^2(k),
\end{gather}\end{subequations}
[Eqs. \eqref{Vieta_2} are the relations existing between the coefficients and the roots of the algebraic equation $\lambda^2-\gamma \lambda+ \omega_c^2=(\lambda-\lambda_1)(\lambda -\lambda_2)=0$, i.e., the Vieta relations for the roots of such a equation (see, e.g., Ref. \onlinecite{Korn})], and bearing in mind Eqs. \eqref{omega_c_primes}, it is easy to see that 
\begin{subequations}\begin{gather}
\lambda_1'(k)=\lambda_2'(k)=0,\\
\lambda_1''(k)=-\lambda_2''(k)={6b\over \rho_c [\lambda_2(k)-\lambda_1(k)]}.
\end{gather}\end{subequations}
Then we further get 
\begin{widetext}\begin{align}
\chi^{-1} &= 
a + 
{3b\over \pi^2\beta\rho_c}
\int
\left\{{1\over 2\omega_c^2(k)}
+
{\psi[1+\lambda_1(k)/\nu]-\psi[1+\lambda_2(k)/\nu]\over \nu [\lambda_1(k) - \lambda_2(k)]}
\right\}{k^2dk}.
\label{chi_simple}\end{align}\end{widetext}
One can see that the expression obtained here by calculating the Landau potential of the system coincides with Eq. \eqref{eq_chi_first}. Consequently, the low-$T$ asymptotics of Eq. \eqref{chi_simple} in the phonon-like and relaxation limiting cases match with those given in Sec.II.A.1 and Sec.II.A.2 respectively.

\section{Structural nonferroelectric transitions: Accounting for acoustic phonons}

In any system, there always exists a coupling between the order parameter $\eta$ and the strain tensor $u_{ij}$ via the term $\eta^2 u_{ll}$ in the potential energy, i.e., the striction effect. As a result of this coupling, acoustic phonons give a contribution to the thermal anomalies associated with low-$T$ structural phase transitions which may be significant because acoustic phonons are excitations of low energy. So let us improve the above studied model by accounting for the striction effect. Thus the model can describe qualitatively nonferroelectrics phase transitions.

Taking into account the striction effect, the potential energy can be written as
\begin{align}
U=\int\Big(&{a\over 2}\eta^2 + {b\over 4}\eta^4 + {c\over 2}(\nabla\eta)^2+{\alpha\over 2}v\eta^2 + 
{K\over 2}v^2 \Big)
d{\mathbf r},
\label{}\end{align}
where $\alpha$ is the striction coefficient, $K$ the bulk modulus and $v$ the dilatation of the system ($v=u_{ll}$). We shall take into account the anharmonicity within the same approximation than in the preceding section. Thus, this potential energy can be written in Fourier space as
\begin{align}
U=U_0(\eta_0,\epsilon)+{1\over 2}\sum_{\mathbf k\not = 0}{\cal K}_{ij}(k,\eta_0,\epsilon)\xi_{i,\mathbf k}\xi_{j,-\mathbf k} 
\label{}\end{align}
(summation over double indices is implied), where 
\begin{align}
U_0(\eta_0,\epsilon)&={a+\alpha\epsilon\over 2}\eta_0^2 + {b\over 4}\eta_0^4 + {K\over 2}\epsilon^2,
\label{U_stric_eps}\end{align}
with $\epsilon$ being the homogeneous dilatation;
\begin{align}
{\cal K}_{ij}(k,\eta_0,\epsilon)=\begin{pmatrix}
a + 3b\eta_0^2 + \alpha\epsilon + ck^2&i\alpha\eta_0k\\
-i\alpha\eta_0k& \rho v_l^2k^2\\
\end{pmatrix},
\label{}\end{align}
with $\rho$ being the density of the system and $v_l$ the longitudinal velocity of sound; and $\boldsymbol \xi_{\mathbf k}=(\eta_{\mathbf k}, u_{\mathbf k})$, with $u_{\mathbf k}$ being the $\mathbf k$-Fourier component of a longitudinal acoustic displacement. 

The system is now a system of linearly coupled harmonic oscillators. Let us consider that relaxation enters in the dynamics of the order parameter only. If this relaxation is similar to that considered in the preceding section, i.e. if it is characterized by a viscosity coefficient $\widetilde \gamma$, the partition function of the system can be calculated (see Ref. \onlinecite{Weiss}). From this partition function we can write down the Landau potential of the system: 
\begin{widetext}
\begin{align}
\Phi(\eta_0)= U_0(\eta_0,\epsilon)+
{1\over\beta}
\sum_{\mathbf k}
\left\{
\ln{\hbar \beta [\lambda_1(k,\eta_0)\lambda_2(k,\eta_0) \lambda_3(k,\eta_0) \lambda_4(k,\eta_0) ]^{1/2}\over \omega_l(k)}
-\sum_{i=1}^4
\ln \Gamma [1+\lambda_i(k,\eta_0)/\nu]
\right\},
\end{align}
where the set $\{\lambda_i\}$ is such that
\begin{subequations}\begin{align}
\lambda_1+\lambda_2+\lambda_3+\lambda_4&=\gamma,\\
\lambda_1\lambda_2+\lambda_2\lambda_3+
\lambda_3\lambda_4+\lambda_4\lambda_1+
\lambda_1\lambda_3+\lambda_2\lambda_4&=\omega_c^2+\omega_l^2,\\
\lambda_1\lambda_2\lambda_3+\lambda_2\lambda_3\lambda_4
+\lambda_3\lambda_4\lambda_1+
\lambda_1\lambda_3\lambda_4&=\gamma\omega_l^2,\\
\lambda_1\lambda_2\lambda_3\lambda_4&=\omega_c^2\omega_l^2-f,
\end{align}\label{Vieta_4}\end{subequations}
with $\omega_c^2(k,\eta_0) ={\cal K}_{11}(k,\eta_0)/\rho_c$, $\omega_l(k)=v_lk$, and $f(k,\eta_0)=(\alpha\eta_0k)^2/(\rho_c\rho)$. Let us mention that Eqs. \eqref{Vieta_4} are the Vieta relations for the roots of the equation
\begin{align}
\lambda^4 - \gamma \lambda^3 +(\omega_c^2+\omega_l^2)\lambda^2-\gamma \omega_l^2 \lambda+ \omega_c^2\omega_l^2-f=0.
\end{align}
The characteristic frequencies of the system are therefore $-i\lambda_i$.

To our purposes, the value of the macroscopic degree of freedom $\epsilon$ can be calculated by minimizing Eq. \eqref{U_stric_eps}. After doing so we find that 
\begin{align}
U_0(\eta_0)&={a\over 2}\eta_0^2 + {b_0\over 4}\eta_0^4,\\ 
{\mathcal K}_{11}(k,\eta_0)&=a + 3b_1\eta_0^2 + ck^2,
\label{}\end{align}
where $b_0=b - \alpha^2/(2K)$ and $b_1=b-\alpha^2/(6K)$. 

Let us now calculate the generalized susceptibility of the system in the symmetric phase ($\eta_0=0$) following the exposed in Sec. II.B. Expressions for the derivates of the $\lambda$'s with respect to $\eta_0$ are then required. From Eqs. \eqref{Vieta_4} one can easily see that $\lambda_i'(k)=0$. For $\lambda_i''(k)$ one finds that, for instance, 
\begin{align}
\lambda_1''(k)
=-{1
\over \lambda_1(k) -\lambda_2(k)}\left(2\omega_c(k)\omega_c''(k)
-{f''(k)\over 
\lambda_1^2(k)+\omega_l^2(k)}\right),
\end{align}
where it has been taken into account that $\lambda_{3,4}(k)=\pm i\omega_l(k)$. Similar expressions for $\lambda_{2,3,4}''(k)$ are also obtainable from Eqs. \eqref{Vieta_4}. 
Moreover the roots $\lambda_{1,2}(k)$ satisfy the relations \eqref{Vieta_2}. Bearing these relationships in mind we find that the inverse of the susceptibility, in addition to Eq. \eqref{suscep_simple},\cite{nota1} has now the contribution
\begin{align}
\Delta\chi^{-1}&=
-{1\over 4\pi^3\beta}
\int{f''(k)\over \omega_c^2(k)}
\left[
{1\over \omega_l^2(k)}
+{2\omega_c^2(k)\over \nu[\lambda_1(k) - \lambda_2(k)]}
\left({\psi[1+\lambda_1(k)/\nu]\over \lambda_1^2(k) + \omega_l^2(k)}
-{\psi[1+\lambda_2(k)/\nu]\over \lambda_2^2(k) + \omega_l^2(k)}
\right)
\right.\nonumber \\ 
&\qquad\qquad\qquad\qquad\quad\left.
+{2\omega_c^2(k)\over \nu \omega_l(k)}
\text{ Im}\left(
{\psi[1+i\omega_l(k)/\nu]\over[\lambda_1(k)-i\omega_l(k)][\lambda_2(k)-i\omega_l(k)]}
\right)\right]k^2dk.
\label{delta_chi_striction}\end{align}

In the purely phononic case, $\gamma=0$ ($\lambda_1=\lambda_2^*=i\omega_c$), Eq. \eqref{delta_chi_striction} reduces to the already known result\cite{Rechester71,Khmelnitskii71}
\begin{align}
\Delta\chi^{-1}&=
{\hbar \alpha^2\over (2\pi)^2\rho_c\rho}
\int
\left(
{n[\omega_c(k)]+1/2\over \omega_c(k)[\omega_c^2(k)-\omega_l^2(k)]}
-
{n[\omega_l(k)]+1/2\over \omega_l(k)[\omega_c^2(k) -\omega_l^2(k) ]}
\right)k^4dk.
\end{align}
\end{widetext}
This contribution yields a temperature dependence of the inverse of the susceptibility $\propto T^4$. This is the leading one far from the transition point [$k_BT\ll \hbar\omega_c(0)$]. Close to the transition [$k_B T\gg \hbar\omega_c(0)$], however, the most important contribution $\propto T^2$ is obtained from Eq. \eqref{suscep_simple}.

Substituting the asymptotic expansion of the psi function into Eq. \eqref{delta_chi_striction} we get its low-$T$ behavior in the relaxation case ($\gamma \not = 0$, $\rho_c\to 0$):
\begin{align}
\Delta\chi^{-1} &
\approx \Delta a^* + \Delta C T^2,
\label{}\end{align}
where $\Delta a^* $ is a constant and $\Delta C = -\{[\alpha^2/[3b_1(K+4\mu/3)]\}C$ [see Eq. \eqref{chi_simple_asym}]. In this case, the importance of acoustic phonons is diminished. This is natural because, due to the relaxation, the dynamics of the order parameter can be activated even far from the transition point. 

\section{Piezoelectric effect}

Let us now study low-$T$ phase transitions taking place in KDP-type systems. In addition to the striction effect, a linear coupling between the order parameter and one of the component of the strain tensor is present in these systems, i.e., a piezoelectric effect. The role that this piezoelectric effect plays in the corresponding transition can be revealed considering the potential energy 
\begin{align}
U=\int\Big(&{a\over 2}\eta^2 + {b\over 4}\eta^4 + {c\over 2}(\nabla\eta)^2+{\alpha\over 2}u_{ll}\eta^2 + du_{xy}\eta \nonumber \\
&+ {\lambda\over 2}u_{ll}^2+ \mu u_{ij}\Big)
d{\mathbf r},
\label{}\end{align}
where $d$ is the piezoelectric coefficient, $\mu$ the shear modulus and $\lambda=K + 2\mu/3$ (for the sake of symplicity, the elastic anisotropy has been taken into account partially). Within the same approximation in the anharmonicity that in preceding sections, in Fourier space this can be written as 
\begin{align}
U=U_0(\eta_0,\hat\epsilon)+{1\over 2}\sum_{\mathbf k\not = 0}{\cal K}_{ij}({\mathbf k},\eta_0,\hat\epsilon)\xi_{i,\mathbf k}\xi_{j,-\mathbf k}, 
\label{U_kdp}\end{align}
where 
\begin{align}
U_0
&={a+\alpha\epsilon_{ll}\over 2}\eta_0^2 + {b\over 4}\eta_0^4 + d\eta_0\epsilon_{xy}+ {\lambda\over 2}\epsilon_{ll}^2+\mu \epsilon_{ij}^2,
\label{U_piezo_eps}\end{align} 
with $\hat\epsilon$ representing the tensor of homogeneous strains;
\begin{subequations}
\begin{align}{\cal K}_{00} &= a + 3b\eta_0^2 + \alpha\epsilon_{ll} + ck^2,\\
{\cal K}_{01} &={\cal K}_{10}^*=i(\alpha\eta_0k_x+{d}k_y/2),\\
{\cal K}_{02} &={\cal K}_{20}^*=i(\alpha\eta_0k_y+{d}k_x/2),\\
{\cal K}_{03} &= {\cal K}_{30}^*=i\alpha\eta_0k_z,\\
{\cal K}_{i,j}&=
\rho [v_t^2 k^2\delta_{ij}+(v_l^2 - v_t^2)k_ik_j]\quad(i,j=1,2,3),
\label{}\end{align}\end{subequations}
with $v_t$ being the transversal velocity of sound; and $\boldsymbol \xi_{\mathbf k}=(\eta_{\mathbf k}, \mathbf{u}_{\mathbf k})$, with $\mathbf u_{\mathbf k}$ being the $\mathbf k$-Fourier component of the displacement vector. Dipolar long-range interactions that also take place in these (ferroelectrics) systems have been neglected. The possibility of such a neglection is explained below.

Let us assume, as in previous section, that if were no coupling between the order parameter and the elastic degrees of freedom ($\alpha=d=0$), the relaxation would be present in the order-parameter dynamics only. This relaxation being characterized by a viscosity coefficient $\widetilde \gamma$, the Landau potential can be written as 
\begin{widetext}
\begin{align}
\Phi(\eta_0)=U_0
+{1\over\beta}
\sum_{\mathbf k}
&\left\{
\ln{\hbar \beta[\lambda_1(k,\eta_0)\lambda_2(k,\eta_0) \lambda_3(k,\eta_0) \lambda_4(k,\eta_0) \lambda_5(k,\eta_0) \lambda_6(k,\eta_0)]^{1/2}\over \omega_l(k)\omega_t(k)}
-\sum_{i=1}^6
\ln \Gamma [1+\lambda _i(k,\eta_0)/\nu]
\right\},
\label{Landau_KDP}\end{align}
where the set $\{\lambda_i\}$ corresponds to the roots of the equation
\begin{align}
\lambda^6-\gamma \lambda^5+(\omega_c^2+\omega_l^2+\omega_t^2)\lambda^4
-(\omega_l^2+\omega_t^2)\gamma \lambda^3
+[\omega_c^2(\omega_l^2+\omega_t^2)+\omega_l^2\omega_t^2+g_l+2g_t]\lambda^2&
\nonumber \\
-\omega_l^2\omega_t^2\gamma \lambda
+\omega_c^2\omega_l^2\omega_t^2+g+g_l\omega _l^2+2g_t\omega_t^2&=0,
\label{ec_sec_6}\end{align}
\end{widetext}
with $\omega_c^2= {\cal K}_{00}/\rho_c$, $\omega_l= v_l k$, $\omega_t= v_t k$; and
\begin{subequations}\begin{align}
g
&={d(v_l^2-v_t^2)\over \rho_c\rho}
(\alpha\eta_0k^2+dk_xk_y) k_xk_y,\\
g_l
&=-{d\over \rho_c\rho}[\alpha\eta_0k_xk_y+d(k_x^2+k_y^2)/4],\\
2g_t
&=-{\alpha\eta_0 \over \rho_c\rho}(\alpha\eta_0k^2+dk_xk_y).
\end{align}\end{subequations}

The values of homogeneous strains can be determined by minimizing Eq. \eqref{U_piezo_eps}. After doing so we find that
\begin{align}
U_0&={a+d^2/(4\mu)\over 2}\eta_0^2 + {b_0\over 4}\eta_0^4,\label{U_0_piezo}\\
{\cal K}_{00}&=a+b_1\eta_0^2+ck^2.
\end{align} 
As we have seen in previous sections, zero-point fluctuations yield a renormalization of the coefficient $a$. But note that, due to the piezoelectric effect ($d \not = 0$), this renormalized coefficient $a^*$ does not vanish at the $T=0$ phase transition: it takes the value $a_c^*=-d^2/(4\mu)$. In consequence, the optical frequencies (or inverse of relaxation times) which enter the above expressions do not vanish at the transition point. This point, that must be taken into account in further calculations, has been overlooked by previous authors.

From the expression \eqref{Landau_KDP} for the Landau potential, the temperature dependence of the susceptibility can obtained in a similar way than the exposed in previous sections. After some straightforward but cumbersome calculations [which implies differentiating with respect to $\eta_0$ and the use of the Vieta relations for the roots of Eq. \eqref{ec_sec_6}] one finds that, in the relaxation case, the low-temperature asymptotics of such a dependence is $\propto T^2$ as in the relaxation cases studied previously. Mention that complicate dispersions arising from dipolar long-range interactions do not alter this result. It is because the temperature dependence is obtained irrespective to integration over wavevectors [see, e.g., Eq. \eqref{coefficient_c}]. This justifies the omission of terms related with these interactions in Eq. \eqref{U_kdp}. 

The fact that the long-range interactions do not change the low-temperature asymptotic of the temperature dependence of the susceptibility in the relaxation case is because of the following. The elementary excitations of lowest energy are that associated with the critical degrees of freedom due to the corresponding relaxation. It is the case whatever the distance from the transition point be, the thermal activation being almost independent on this distance. 

If there is no relaxation, the elementary excitations of lowest energy are, however, the acoustic ones. We have already seen in Sec. III a similar situation that takes place in nonferroelectrics: (longitudinal) acoustic phonons yield the main contribution far from the transition point. What is different now is that acoustic phonons may yield the main contribution even at the transition point. It is because, as we have already noticed, there is no vanishing optical frequencies due to the piezoelectric effect. Estimations of the corresponding gap show that in KDP, for example, optical phonons could be frozen out up to temperatures of $\sim 10~\rm K$.\cite{Cano03} We shall consider in further calculations that this freezing actually takes place, what seems to be well possible many other systems (the piezoelectric effect in KDP is not especially strong). Then for the low-$T$ asymptotics of the temperature dependence of the inverse of the susceptibility one finds that
\begin{align}
\chi^{-1} \approx \begin{cases}
a'+(T/\Theta_1)^{5/2},&\displaystyle a'\ll {k_B T \over \hbar v_t}
(c |a_c^*|)^{1/2}, \\ \\
a'+(T/\Theta_2)^4,&\displaystyle a'\gg {k_B T \over \hbar v_t}
(c |a_c^*|)^{1/2},
\end{cases}\end{align}
where $a'=a^*-a_c^*$ and $\Theta_{1,2}$ are constants.\cite{nota2} Due to the combined influence of both striction and piezoelectric effects, the $\sim T^{5/2}$ behavior close to the transition point is obtained. These two effect has not been considered both together until now, so the above mentioned asymptotics has been overlooked. Mention that it is related to the softeing that, due to the piezoelectric effect, there exists in an acoustic branch, i.e., vanishing velocity of a transversal acoustic wave. Dipolar interactions do not modify drastically such a softening, what justifies their omission in Eq. \eqref{U_kdp}. 

\section{Conclusions}

We have presented a semiphenomenological theory of low-$T$ structural phase transitions by including order-parameter dynamics of relaxation type. We have restricted ourselves to the case in which this relaxation does not depend on neither the wavevector nor the frequency. Within the continuous media theory we use, the relaxation is then characterized by a single phenomenological parameter (viscosity coefficient). Both phonon-like and relaxation limiting cases can be reproduced by varying this parameter. 

We have focused our attention in the low-$T$ asymptotics of the temperature dependence of the generalized susceptibility. A dependence $\sim (T^2-T_c^2)^{-1}$ close to the transition point is commonly accepted as an indication of a displacive transition, i.e. a transition involving phonon-like order-parameter dynamics, although the presence of long-range interaction modifies such a asymptotics.\cite{Kvyatkovskii01} In particular, we have found that due to the combined effect of striction and piezoelectric effects (KDP case) this asymptotics becomes into $\sim (T^{5/2}-T_c^{5/2})^{-1}$. These two effects have not been considered both together until now, so this asymptotics is overlooked in previous papers. For order-disorder transitions (relaxational dynamics), we have found that this dependence is $\sim (T^2-T_c^2)^{-1}$ for temperature-independent relaxation times, irrespective to the long-range interactions. So, in principle, in KDP-type systems low-$T$ structural phase transitions with phonon-like and relaxational order-parameter, i.e. displacive and order-disorder respectively, might be distinguishable by these asymptotics; although experimentally it is quite difficult. Let us stress that the key point for understanding the low-$T$ properties is the order-parameter dynamics, what could be determined in ferroelectrics by measuring e.g. the corresponding dielectric losses.


\begin{thebibliography}{0}
\bibitem{nota0} It is quite probable that, due to the large masses of the ordering units usually involved in structural order-disorder transitions (not to mention decoherence), to convert these transitions into displacive ones factors such as quantum tunneling may be effective only at very low temperatures not available in experiments.
\bibitem{nota1} The constant $b$ has now to be substituted by $b_1$ in Eq. \eqref{suscep_simple}.
\bibitem{nota2} $\eta$ representing polarization, and using CGS units, $\Theta_{1,2}$ can be estimated as Debye temperatures.
\end{thebibliography}
\end{document}